\documentclass[aps,prb,groupedaddress,twocolumn]{revtex4}

\usepackage{amsmath}
\usepackage{graphicx}
\usepackage{dcolumn}
\usepackage{bm}
\begin{document}

\title{Vortex pinning by meandering line defects in planar superconductors}

\author{Eleni Katifori}\email{katifori@fas.harvard.edu}
\author{David R. Nelson}
\affiliation{Department of Physics, Harvard University, Cambridge, Massachusetts, 02138}

\date{\today}

\begin{abstract}

To better understand vortex pinning in thin superconducting slabs, we study 
the interaction of a single fluctuating vortex filament with a curved line defect in 
(1+1) dimensions. This problem is also relevant to  the interaction of scratches with wandering 
step edges in vicinal surfaces. The equilibrium probability density for a fluctuating line attracted 
to a particular fixed defect trajectory is derived analytically by mapping the problem to a straight line defect in the presence 
of a space and time-varying external tilt field. The consequences of both rapid and slow changes in the frozen defect trajectory, 
as well as finite size effects are discussed. A sudden change in the defect direction leads to a delocalization transition, accompanied by a 
divergence in the trapping length, near a critical angle.

\end{abstract}
\maketitle

\section{\label{sec:introd}Introduction}

Understanding the physics of thermally wandering lines interacting with extended defects is crucial 
in explaining the properties of  planar high Tc superconductors permeated
by columnar pins or twin boundaries\cite{blatt94}. Considerable progress is possible theoretically in planar superconductors,
when platelet samples are permeated by a single sheet of vortex lines due to a small in-plane
magnetic field \cite{ hofstett04, affl04}. Related problems arise on vicinal surfaces, where thermally fluctuating step edges \cite{jeong99} can interact 
with scratches, grain boundaries or terraces created by lithography. In superconductors, a zoo of different systems whose properties 
are determined by the interplay of the interacting vortex filaments, thermal fluctuations and pinning 
has been thoroughly explored when the attracting defect is straight or vortices follow trajectories 
determined by point disorder \cite{blatt94,hatan97, hofstett04, fish91, hwa93b, derev94}. However, little work 
has been done so far when the defects themselves follow a controlled but non-trivial trajectory. On vicinal surfaces,
such defects could be created by etching a wavy, semipermanent scratch, or studied in the context 
of the square terraces created lithographically by Lee and Blakely \cite{blakely99}.
One might also consider a thin high Tc superconductor sample where a wavy notch has been etched, although 
curved extended defects can arise naturally in the form of grain boundaries in polycrystalline
platelet superconductors \footnote{E. Zeldov, private communication.}. Finally, close intersection of splayed
columnar defects \cite{hwa93,dous94} provide preferred vortex tracks with a sudden change of direction.

In this work we investigate a single thermally fluctuating 
line that interacts with a single quenched meandering defect. This situation mimics the dilute flux line limit of a planar 
superconductor with a single pinning defect when the vortices are far apart enough to be considered noninteracting. Although
we use the terminology of flux lines throughout, similar results should apply to scratch-step interactions on vicinal surfaces. The case 
of many step edges (or vortex lines) interacting with a single curved line defect will be treated in a future publication \footnote{E. Katifori and D. R. Nelson, to be published.}. 

\begin{figure}
\includegraphics[scale=0.5]{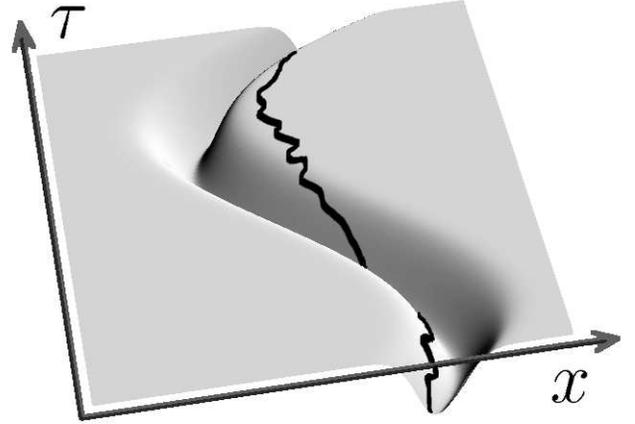}
\caption{\label{fig:figure1}Thermally excited fluctuating string (vortex line or step edge) interacting 
with a meandering attractive line-like potential. The dark line indicates the elastic string 
trajectory $x(\tau)$, the energy of which is given by Eq.(\ref{ener}).}
\end{figure}

In the spirit of the treatment of high Tc superconductors in Ref.\cite{hatan97}, 
we describe the trajectory of a flux line in (1+1) dimensions as a classical elastic string $x(\tau)$ 
subject to thermal fluctuations. We assume high enough temperatures and sufficiently clean samples so that 
point disorder can be neglected. In a quantum analogy \cite{ hofstett04, affl04}, (see below), the spatial direction 
labelled by $\tau$ plays the role of imaginary time.
We assume that overhangs are improbable, so the choice of a single valued
function to describe a string-like flux line is adequate. The trajectory of the defect itself (a notch, etched on the thin superconducting slab, say)
will be represented as $x_o(\tau)$,
and attracts the vortex with a short range attractive potential $V[x-x_o(\tau)]$, (see Fig.\ \ref{fig:figure1}).

For the energy of a flux line induced by a magnetic field along $\tau$ in (1+1) dimensions, bounded by time-like initial 
and final coordinates $\tau_i$ and $\tau_f$, we take the integral: 
\begin{equation}\label{ener}
E[x(\tau)]=\int\limits_{\tau_i}^{\tau_f}d\tau \left[ \frac{\gamma}{2}\left(\frac{dx(\tau)}{d\tau}\right)^2+V[x(\tau)-x_o(\tau)]\right],
\end{equation}
where $\gamma$ is the coarse-grained line tension. The finite temperature behavior of the string can be determined from the partition
function, which is a functional integral over all possible vortex configurations $x(\tau)$, 
\begin{equation}\label{part}
\mathcal{Z}=\int \mathcal{D} x(\tau) e^{-E[x(\tau)]/T}
\end{equation}
where we have set Boltzmann's constant, $k_B\equiv 1$.

The statistical physics of the above system is controlled by a competition between the two terms of the line energy of Eq.(\ref{ener}). 
The ``kinetic energy'' (corresponding to the line tension of the string) is minimized when
the vortex line is straight and parallel to the $\tau$ direction, whereas the ``potential energy'' is minimized when the string resides at the deepest point of 
$V[x(\tau)-x_o(\tau)]$, namely exactly on the attracting defect. Depending on the strength of the potential, the stiffness
of the string, the temperature and the trajectory $x_o(\tau)$ of the defect, the elastic string will either localize near the trajectory of the
defect or wander far from it. Since in most experimental realizations the attractive potential is relatively short ranged, in
this work  we will take $V[x]$ in Eq.(\ref{ener}) to be an attractive delta function, which should adequately describe the long wavelength properties of the system.

The case of $x_o(\tau)$ being a zeroth or first order polynomial of $\tau$ (i.e., a straight line!) has been studied in detail 
in Ref.\ \cite{hatan97}, with emphasis on the delocalization transition that takes place for large relative tilts of the pining
defect and the flux line. This work adapts the results and conclusions of that paper to the case of a ``bent defect'' with a sudden change in direction,
and considers more general quenched defect trajectories as well.

This paper will be structured as follows. In Section \ref{sec:propag} we review the basic mapping from a classical 
description of the thermally fluctuating flux line to a (non-Hermitian) quantum mechanical problem, and use this mapping to 
analytically compute the partition function (i.e., the propagator for the equivalent quantum problem) for the straight defect case. 
Section \ref{sec:probabil} explains how to use this propagator to derive exact 
expressions for a simple realization of a ``bent'' defect trajectory and determine how the defect probability distribution responds to the presence of the kink.
Section \ref{sec:adiabexp} describes a perturbative solution
in the ``Born-Oppenheimer'' limit of a slowly varying $x_o(\tau)$.
Finally, Appendix \ref{sec:compdet} contains some 
calculational details, relevant to Sections \ref{sec:propag} and  \ref{sec:probabil}.

\section{\label{sec:propag}Partition function for a flux line interacting with a linear defect}

In this section we briefly review the connection between the classical path integral of Eq.(\ref{part})
and the ``quantum mechanical'' formulation, which follows from a transfer matrix technique \cite{Feyn}. We then proceed to 
analytically compute the partition function that governs
the properties of the thermally excited vortex line interacting with a straight line defect.

For clarity we denote the partition function of Eq.(\ref{part}) as $\mathcal{Z}[x_f,\tau_f;x_i,\tau_i;x_o(\tau)]$, 
where $x_i$ and $x_f$ are fixed initial and final positions of the flux line. (We may eventually choose to integrate over these endpoints.)
Since the potential is only implicitly dependent on $\tau$ through the defect coordinate $x_o(\tau)$, we can map the problem by a
change of variables  $x(\tau)\rightarrow y(\tau)=x(\tau)-x_o(\tau)$ onto one of a straight defect in a $\tau$-dependent external tilt field  
$h(\tau)=dx_o(\tau)/d\tau$ that imposes tendency to drift on the flux line. The Jacobian of the transformation is unity, so the measure of the path integral remains the same.
After this transformation, the partition function reads:
\begin{equation}\label{parttrans}
 \mathcal{Z}_t[\small x_f-x_o(\tau_f),\tau_f;x_i-x_o(\tau_i),\tau_i;h(\tau)]=\int\mathcal{D} y(\tau) e^{-E[y(\tau)]/T}
\end{equation}
where
\begin{equation}\label{ener2}
E[y(\tau)]=\int\limits_{\tau_i}^{\tau_f}d\tau \left[ \frac{\gamma}{2}\left(\frac{dy(\tau)}{d\tau}+h(\tau)\right)^2-V_o\:\delta (y)\right].
\end{equation}
The sign of $V_o>0$ is chosen so that the potential is attractive.
By construction, the old and new partition functions must agree: 
\begin{equation}\label{equivl}
\mathcal{Z}[x_f,\tau_f;x_i,\tau_i;x_o(\tau)]=\mathcal{Z}_t[x_f-x_o(\tau_f),\tau_f;x_i-x_o(\tau_i),\tau_i;h(\tau)].
\end{equation}
One way around the technical difficulties (see, e.g., Ref.\cite{klein95}) of the evaluation of the partition function of Eq.(\ref{parttrans}) 
due to the singular nature of the delta function, is to 
recast the problem in quantum language. The classical partition function then becomes a matrix element of the imaginary time evolution operator:
\begin{equation}\label{part2}
\mathcal{Z}_t[x_f,\tau_f;x_i,\tau_i;h(\tau)]=\langle x_f|T_{\tau}\{e^{-\int_{\tau_i}^{\tau_f}d\tau \mathcal{H}(\tau)/T}\}|x_i \rangle,
\end{equation}
where the time ordering operator $T_{\tau}$ is required by the time-dependent Hamiltonian:
\begin{equation}\label{hami}
\mathcal{H}(\tau)= -\frac{T^2}{2\gamma}\:\frac{\partial^2}{\partial x^2}-h(\tau)\:T\:\frac{\partial}{\partial x}-V_o\:\delta(x).
\end{equation}
The mapping onto imaginary time quantum mechanics is clear if we let $T\rightarrow\hbar$ and $\gamma\rightarrow m$, where $m$
is a particle mass. The term $h(\tau)\:T\:\frac{\partial}{\partial x}$ 
makes the Hamiltonian non-Hermitian.

Before considering curved defect trajectories, we analytically compute the matrix element $\mathcal{Z}_t[x,\tau;y,0;h]$ for constant tilt $h$, by expanding 
with respect to the complete set of right and left energy eigenstates $\{|\psi_K\rangle_R\}$ and $\{{}_L\langle \psi_K|\}$  of the non-Hermitian Hamiltonian 
of Eq.(\ref{hami}):
\begin{equation}\label{expa}
\mathcal{Z}_t[x,\tau;y,0;h]=\sum\limits_K \langle x|\psi_K\rangle_R e^{-E_K\: \tau/T}{}_L\langle\psi_K|y\rangle,
\end{equation}
with $E_K$ being the $K^{\mathrm{th}}$ eigenenergy. Since both 
the Hamiltonian and the propagator are $\tau$-translationally invariant, without loss of generality we have chosen zero as the origin of imaginary time. For 
notational simplicity we have also rescaled the spatial coordinates ($x$,$\tau$) and tilt 
$h$ so that $[x]= T^2/\gamma V_o$, $[\tau]=2 T^3/\gamma V_o^2$,  $[h]=V_o/T$. After these
rescalings, the evolution equation for $\mathcal{Z}_t[x,\tau;x_i,\tau_i;h]\equiv\mathcal{Z}[x,\tau]$ reads:
\begin{eqnarray}
\frac{\partial \mathcal{Z}[x,\tau]}{\partial{\tau}}&=&-\mathcal{H}\mathcal{Z}[x,\tau]\nonumber\\
&=&\left(\frac{\partial^2}{\partial x^2}+2\: h(\tau)\frac{\partial}{\partial x}+2\delta(x)\right)\mathcal{Z}[x,\tau].
\end{eqnarray}
As explained in Ref.\cite{hatan97}, there is a critical value $h_c$ of the tilt, above which the flux line delocalizes from the defect. 
Our rescaled units are chosen so that this value is $h_c=1$ for our delta function potential. Notice that in the rescaled units
the track of the straight line defect is given by $\frac{dx_o(\tau)}{dt}=2h$.

The set of normalizable eigenstates of Eq.(\ref{hami}) is different above and below the critical tilt 
$h_c$ (see Appendix \ref{sec:compdet}); however, the propagator when 
the lateral dimension $L_x$ of the sample tends to infinity has the same form for both regimes, namely:
\begin{multline}\label{part3}
\mathcal{Z}_t[x,\tau;y,0;h]=e^{-h(x-y)-h^2\tau}\bigg[\frac{e^{-(x-y)^2/4\tau}}{2\sqrt{\pi\tau}}+\\
+\frac{e^{-|y|-|x|+\tau}}{2}\mathrm{erfc}\left(\frac{|y|+|x|}{2\sqrt{\tau}}-\sqrt{\tau}\right)\bigg]
\end{multline}
where $\mathrm{erfc}(x)=\frac{2}{\sqrt{\pi}}\int\limits_x^{\infty}dt\: e^{-t^2}$.
The first term inside the square brackets is independent of the defect, translationally invariant in the $x$ direction and describes 
the random thermal ``diffusion with drift'' of the flux line, $\mathcal{Z}\sim \exp[-(x+2 h \tau)^2/4\tau]$
\footnote{Returning to the original, tilted defect frame of reference, we see that $\mathcal{Z}\sim \exp[-x^2/4\tau]$
as expected from Eq.(\ref{hami2}) for $V_o\rightarrow 0$.} . This term dominates for small times and/or large $|x|$, $|y|$, i.e. far from
the defect.
The second term becomes important for sufficiently large imaginary times or small $|x|$, $|y|$. The delocalization transition
manifests itself in the exponential growth of the second term for large $|x|$, $|y|$ when the tilt $h$ in the prefactor exceeds the critical tilt $h_c=1$.
Upon returning to the original units of $x$ and $\tau$, we see that the effect of the defect becomes apparent for $\tau>\tau_c^*\simeq T^3/\gamma V_o^2$,
provided $|x|$ and $|y|$ are within a localization length $x^*=T^2/\gamma V_o$ of the defect at the origin. For $V_o\rightarrow 0$, the first term in 
the brackets scales as $1/V_o$ and will dominate over the second term which becomes independent of $V_o$. This limit leads to the expected 
free flux line propagator. These results can be easily translated in the original frame of reference (where the defect is tilted) by the use of Eq.(\ref{equivl}).

\section{\label{sec:probabil}PROBABILITY DISTRIBUTION OF THE FLUCTUATING vortex filament}

\begin{figure}
\includegraphics[scale=0.5]{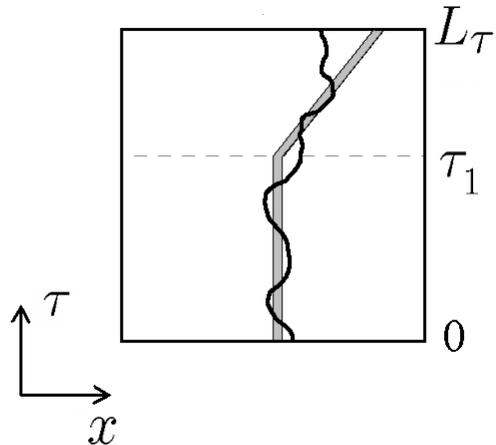}
\caption{\label{fig:figure2}Flux line near a kinked defect: The defect, initially parallel to the average direction of the freely 
fluctuating vortex suddenly tilts at $\tau=\tau_1$.}
\end{figure}

In this section we use the above propagator to analytically compute the probability distribution function of a flux line
interacting with a $\tau$-dependent defect. As a simple, yet revealing example we study a line defect consisting
of two straight lines joined at an angle, at $\tau=\tau_1$, namely: $x_o(\tau)=2h\: (\tau-\tau_1)\Theta(\tau-\tau_1)$, see Fig.\ \ref{fig:figure2}. 
Here, $\Theta(x)$ is the step function, $\Theta(x)=0$, $x<0$, $\Theta(x)=1$, $x>0$. Any
sufficiently well behaved defect trajectory can be approximated by a series of finite segments; the method 
described below can be straightforwardly generalized to more complicated piece-wise linear configurations.

When $h<1$, and in particular when $h=0$  for the first segment of the defect, the system relaxes to the localized ground state exponentially 
fast as $\tau$ becomes large. A ground state initial condition $|\psi^i\rangle=|\psi_g\rangle$ at $\tau=0$ is thus the experimentally relevant case 
when the kink at $\tau_1$ is sufficiently far  from the boundaries. For $\tau=L_{\tau}$ we choose free boundary 
conditions embodied in the final state$\langle\psi^f|=\int dy \langle y|$, meaning that no restriction is imposed on the exit point of the vortex from the 
sample. In what follows, unless otherwise stated, we work in the thermodynamic limit in the $x$ direction, so integrations
such as $\int dx$ may be understood as $\int_{-\infty}^{\infty}dx$.

The probability distribution for the flux line on some internal time slice $\tau$ is formally equal to:
\begin{equation}\label{prob}
P(x,\tau)=\frac{1}{\mathcal{Z}}  \langle\psi^f| S(L_{\tau},\tau) |x\rangle  \langle x| S(\tau,0) |\psi^i\rangle,
\end{equation}
where 
\begin{equation}\label{Stran}
S(\tau_f,\tau_i)\equiv T_{\tau}\{e^{-\int_{\tau_i}^{\tau_f}d\tau\;\mathcal{H}(\tau)/T}\}
\end{equation}
is the time ordered evolution operator associated with the Hamiltonian
\begin{equation}\label{hami2}
\mathcal{H}(\tau)=-\frac{T^2}{2\gamma}\frac{\partial^2}{\partial x^2}-V_o\:\delta[x-x_o(\tau)].
\end{equation}
In Eq.(\ref{prob}), $\mathcal{Z}$ is used to represent the partition function for the entire length of the system on 
both sides of the kink,
\begin{equation}\label{part4}
\mathcal{Z}\equiv\langle\psi^f| S(L_{\tau},0) |\psi^i\rangle. 
\end{equation}

Since the defect trajectory $x_o(\tau)$ is piecewise straight, we can split the factors in the numerator 
of Eq.(\ref{prob}) into two pieces such that $dx_o/d\tau=const$. For computational 
purposes, the cases $\tau>\tau_1$ and $\tau<\tau_1$ need to be treated separately. 

First, for $\tau<\tau_1$ we expand the probability distribution in position eigenstates as:
\begin{widetext}
\begin{eqnarray}\label{prob2}
P(x,\tau)&=&\frac{1}{\mathcal{Z}}\iint \mathrm{d}y \mathrm{d}y' \langle y |S(L_{\tau},\tau_1)|y'\rangle \langle y'|S(\tau_1,\tau)|x\rangle \langle x |S(\tau,0)|\psi_g\rangle\nonumber\\
&=&\frac{1}{\mathcal{Z}}\iint \mathrm{d}y \mathrm{d}y'  \mathcal{Z}[y,L_{\tau};y',\tau_1;x_o(\tau)]  \mathcal{Z}[y',\tau_1;x,\tau;0] \langle x |S(\tau,0)|\psi_g\rangle
\end{eqnarray}
\end{widetext}
The last propagator inside the integrand effectively localizes the vortex near the defect at $x_o(\tau)=0$ case. 
Since the initial condition is the ground state of 
the time independent Hamiltonian, this last factor is just equal to $e^{-|x|}e^{-E_g\tau}=e^{-|x|}e^{\tau}$, 
since the ground state energy in our units is $E_g=-1$.
 The middle propagator in Eq.(\ref{prob2}) follows from Eq.(\ref{part3}) by setting $h=0$. 
The leftmost factor in the integrand can also be expressed in terms of  
Eq.(\ref{part3}), if we first perform the tranformation of Eq.(\ref{equivl}). 
Then, since $y'$ is a dummy integration variable and $x_o(\tau_1)=0$, it is easy to see that Eq.(\ref{prob2}) can be rewritten in terms of the transformed 
partition function appearing in Eq.(\ref{equivl}) as:
\begin{multline}\label{probl}
P(x,\tau)=\frac{1}{\mathcal{Z}}\iint \mathrm{d}y\mathrm{d}y'  \mathcal{Z}_t[y,L_{\tau};y',\tau_1;h]\cdot\\  
\cdot\mathcal{Z}_t[y',\tau_1;x,\tau;0]e^{-|x|}e^{\tau},\:\:\:\:(\tau<\tau_1).
\end{multline}
Although the $y$-integration can be computed analytically (see Appendix \ref{sec:compdet}), we have resorted to numerical integration to handle the integration over $y'$.

Upon following a similar procedure for $\tau>\tau_1$, we obtain:

\begin{multline}\label{problla}
P(x,\tau)=\frac{1}{\mathcal{Z}}\int \mathrm{d}y  \mathcal{Z}_t[y,L_{\tau};x-x_o(\tau),\tau;h]\cdot \\
\cdot\int \mathrm{d}y' \mathcal{Z}_t[x-x_o(\tau),\tau;y',\tau_1;h]e^{-|y'|}e^{\tau_1}, \\
(\tau>\tau_1).
\end{multline}
As shown in Appendix \ref{sec:compdet}, in this case both the integrals can be done analytically.

\begin{figure}
\includegraphics[scale=0.4]{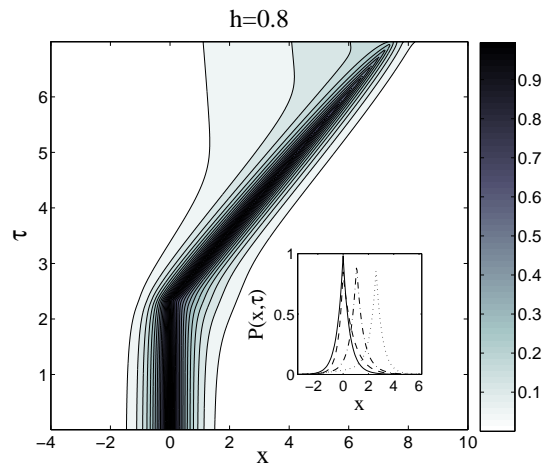}
\caption{\label{fig:allg08}Contour plot for the probability density distribution of the line defect for a scratch everywhere below the critical tilt angle: 
pinned regime. The darker areas correspond to higher probability.
The tilt of the upper segment of the defect is $h=0.8<h_c=1$, the cusp is at $\tau_1=7/3\approx 2.33$ 
and the total length of the sample is $\L_{\tau}=7$. The inset
shows the probability density distribution for: $\tau=1$ solid line, $\tau=2.33$ dashed line, $\tau=3$ dash-dotted line, $\tau=4$ dotted line. 
Note that the distribution, while remaining localized,
develops an asymmetry for $\tau>7/3$.}
\end{figure}

\begin{figure}
\includegraphics[scale=0.4]{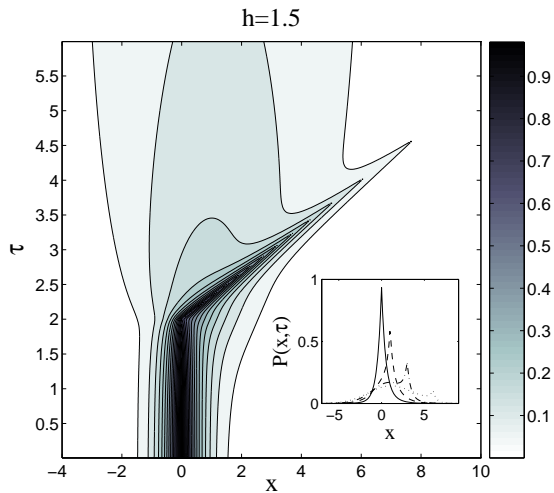}
\caption{\label{fig:allg15}Contour plot for the probability density distribution of the defect line 
whose upper part is above the critical tilt angle for the scratch: depinned regime. 
The tilt of the upper segment of the defect now is $h=1.5>h_c=1$, while $\tau_1=2$ and $\L_{\tau}=6$. The inset shows
the probability density distribution for: $\tau=1$ solid line, $\tau=2.33$ dashed line, $\tau=3$ dash-dotted line, 
$\tau=4$ dotted line. Note that the probability distribution begins to delocalize for $\tau>\tau_1=2$.}
\end{figure}

Having now computed $P(x,\tau)$ for all imaginary times, we can examine the flux line probability distribution for this simple realization of
an attractive line defect that suddenly changes direction. We have studied two cases, one where the tilt of the upper segment of the defect is below 
the critical value $h_c=1$ for delocalization (Fig.\ \ref{fig:allg08}) and one where it is above (Fig.\ \ref{fig:allg15}).

Note the following: For time-like coordinates
below the kink at $\tau_1$,  the probability density does not differ much from the ground state probability density of Ref.\cite{hatan97}
$P(x,\tau)=e^{-2|x|}$. However, as the position $\tau_1$ of the kink in the scratch is approached from below, there is a broadening of the probability density 
on the right or ``concave'' side of the defect, and a similar narrowing on the left or ``convex'' (see Fig.\ \ref{fig:allg08}).
This phenomenon is not directly related to the delocalization of 
the flux line that occurs for tilts greater than the critical tilt, because it is present for both $h>1$ and $h<1$ 
(recall that the critical tilt is $h_c=1$). For $h<1$, after this 
small shifting of the probability density in the vicinity of the kink, the probability again resumes the ground state distribution. 
When $\tau\lesssim L_{\tau}$ we notice a small broadening of the probability which is a boundary condition effect.

When the tilt of the second segment of the defect is above the critical tilt $h_c=1$ (see Fig.\ \ref{fig:allg15}), 
the probability density remains localized around 
the defect only for some small distance above the kink. Two local maxima of probability density appear, one 
localized on the defect and the other representing an approximately gaussian probability distribution. As we move further away from the kink, the 
probability of observing the flux line close to the defect reduces, and the weight associated with the delocalizing diffusive part increases.

\begin{figure}
\includegraphics[scale=0.6]{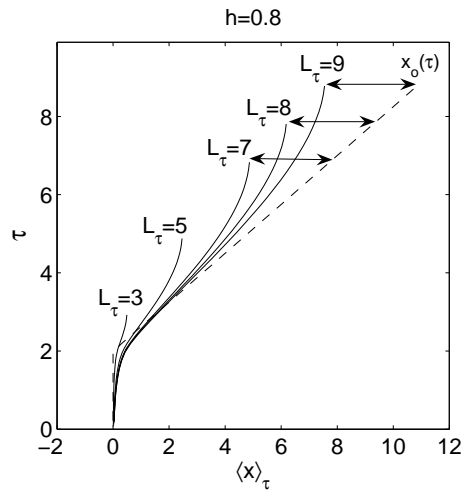}
\caption{\label{fig:figt2g07}Average position of the vortex $\langle x\rangle_{\tau}$ : pinned regime. Here $\tau_1=2$ and $h=0.8$. The dashed line is is the scratch trajectory and
the solid lines represent the thermal average $\langle x \rangle_{\tau}$ for different sample lengths. The arrows mark the average distance of the step from the exit
point of the scratch. As $L_{\tau}$ increases, this distance approaches a constant shift in the trajectory.}
\end{figure}

\begin{figure}
\includegraphics[scale=0.6]{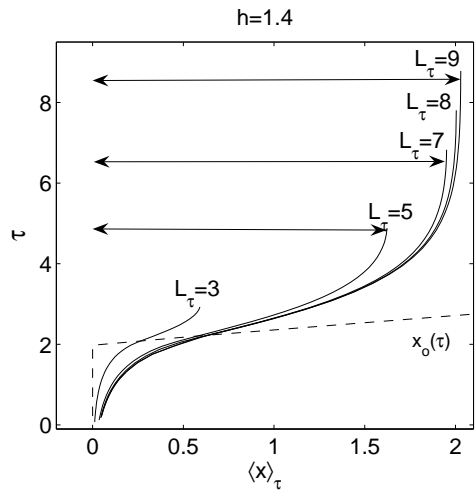}
\caption{\label{fig:figt2g13}Average position of the vortex $\langle x\rangle_{\tau}$: depinned regime. Here $\tau_1$=2 and $h$=1.4. The arrows represent the
deflection of the vortex trajectory due to the tilted scratch and the dashed line is again the defect trajectory. $\langle x \rangle_{L_{\tau}}$ 
converges to $\sim 2.2$ as $L_{\tau}$ increases. }
\end{figure}

\begin{figure}
\includegraphics[scale=0.4]{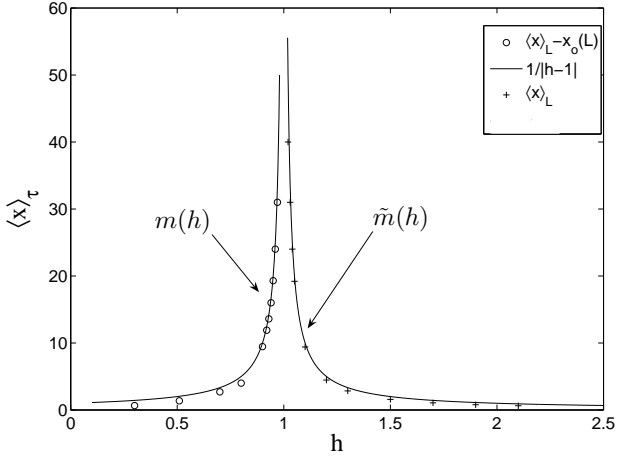}
\caption{\label{fig:depinningtransition}Shift in the line trajectory near the depinning transition; curve to the left of $h_c$=1 represents $m(h)$
while that to the right represents $\tilde{m}(h)$.}
\end{figure}

To better understand the depinning transition as the slope of the kink angle approaches the critical value $h_c$ from below,
we have computed the average position of the fluctuating line $\langle x\rangle_{\tau}=\int dx\:x\:P(x,\tau)$ for several tilts and sample lengths. 
 In Fig.\ \ref{fig:figt2g07} we have plotted the average position of the fluctuating vortex line for
several total lengths in the imaginary time direction $L_{\tau}$. The position of the kink is held fixed at $\tau_1=2$ and the tilt is $h=0.8$. 
We observe that as $L_{\tau}$ grows, the distance $m$ of the exit point of the vortex line $\langle x\rangle_{L_{\tau}}$ 
from the defect, indicated in the figure, tends to a constant.
We have computed
\begin{equation}
 m(h)\equiv\lim\limits_{L_{\tau}\to\infty}(\langle x\rangle_{L_{\tau}}-x_o(L_{\tau})),\:\:\:\:h<1,
\end{equation}
by obtaining values for up to $L_{\tau}=50$ and then extrapolating $L_{\tau}$ to infinity. 
The convergence becomes very slow as $h\to 1^-$. The limiting offset $m(h)$
serves as an indicator of the depinning transition for $h\rightarrow h_c^-$. 
Similarly, for $h>1$, as can be seen from Fig.\ref{fig:figt2g13}, a corresponding indicator is 
the offset of  $\langle x\rangle_{L_{\tau}}$ from the origin, i.e. the original location of the defect. We now measure a shift
\begin{equation}
\tilde{m}(h)\equiv\lim\limits_{L_{\tau}\to\infty}(\langle x\rangle_{L_{\tau}}-x_o(0)),\:\:\:\:h>1,
\end{equation}
which contributes to the total magnetization, 
in the vortex line case.
By a fitting of $m(h)$ and $\tilde{m}(h)$ to $1/|h-1|^{\nu}$ (see Fig.\ \ref{fig:depinningtransition}) we find the critical exponent of the 
delocalization transition to be $\nu=1.0$,\footnote{
$\tilde{m}(h)$ is related to the trapping length introduced in \cite{hatan97} as a measure of the offset 
of the flux line from the average position it would assume 
without the defect (here that would be $\langle x \rangle_{\tau}=0$ because of the boundary condition) and the critical exponent computed here
agrees with the exponent analytically predicted there.}. An identical exponent describes the delocalization transition studied in Ref. \cite{hatan97}.

\

\section{\label{sec:adiabexp}Solution for a random trajectory in the adiabatic limit}

In this section we use the time dependent perturbation theory developped in Mostafazadeh \cite{mostafa97} 
to analytically study the case of a slowly bending defect, i.e. $\left|\frac{dh}{d\tau}\right|\ll 1$.
We will compute the probability distribution defined by Eq.(\ref{prob}), assuming that the initial and final states are the ground state, 
which should be a good approximation for physical systems
with $h=$const for $\tau$ everywhere outside the interval $(0,L_{\tau})$. As in Sec.\ \ref{sec:propag}, we will preform a defect coordinate tranformation 
\begin{equation}
x\rightarrow x+x_o(\tau)
\end{equation}
and reduce the problem to a straight defect in
a time dependent external tilt field.

We will assume that $|h(\tau)|<1$. For any fixed $h(\tau)$ we can then be sure of the existence of a bound state. 
We suppose for the moment that $L_x<\infty$, which allows us to consider a \textit{discrete} spectrum of free states. We can then write the matrix 
element of the time evolution operator as a product over time slices:
\begin{eqnarray}\label{apppro}
&{}&\langle x|S(\tau,0)|\psi_g\rangle=\lim\limits_{N\rightarrow\infty}\sum\limits_{\psi_1,\dots,\psi_N}e^{-\sum\limits_{j=0}^{N}E_{\psi_j}(t_j)\epsilon}\nonumber\\
&{}&\qquad\times\prod\limits_{k=1}^N\langle x|\psi_N(\tau)\rangle_R{}_L\langle \psi_k(t_k)|\psi_{k-1}(t_{k-1})\rangle_R 
\end{eqnarray}
where $\epsilon\equiv\tau/N$, $t_k=k\;\epsilon$ and $|\psi_{k}(t_{k})\rangle_R$ and ${}_L\langle \psi_k(t_k)|$ are respectively the instantaneous right and left eigenstates of 
$\mathcal{H}(\tau)$ for $h=h(\tau=t_N)$, namely:
\begin{equation}
\mathcal{H}(\tau)|\psi_{k}(t_{k})\rangle_R=E_{\psi_k}(t_k)|\psi_{k}(t_{k})\rangle_R.
\end{equation}
Each of the summations embodied in  $\sum\limits_{\psi_1,\psi_2,\dots,\psi_N}$ runs over the entire spectrum, while $|\psi_0(0)\rangle_R\equiv|\psi_g(0)\rangle_R$.

In the limit $\epsilon\rightarrow 0 $, we can expand $|\psi_{k-1}(t_{k}-\epsilon)\rangle_R$ as $|\psi_{k-1}(t_{k}-\epsilon)\rangle_R\simeq |\psi_{k-1}(t_{k})\rangle_R-\epsilon \frac{d}{dt}|\psi_{k-1}(t_{k})\rangle_R$, so:
\begin{equation}\label{appmat}
{}_L\langle \psi_k(t_k)|\psi_{k-1}(t_{k-1})\rangle_R =\delta_{\psi_k,\psi_{k-1}}-\epsilon A_{\psi_k,\psi_{k-1}}(t_k)+O(\epsilon^2),
\end{equation}
where 
\begin{equation}
A_{\psi_k,\psi_{k-1}}(t_k)\equiv \langle \psi_k(t_k)|\frac{d}{dt}|\psi_{k-1}(t_{k-1})\rangle_R.
\end{equation} 
To linear order in $\epsilon$, we can rewrite (\ref{appmat}) as a sum of two mutually excluding terms:
\begin{multline}\label{appmat2}
 {}_L\langle \psi_k(t_k)|\psi_{k-1}(t_{k-1})\rangle_R =e^{-\epsilon A_{\psi_k,\psi_{k}}(t_k)}\big[\delta_{\psi_k,\psi_{k-1}}+\\
+\epsilon e^{\epsilon A_{\psi_k,\psi_{k}}(t_k)}
(\delta_{\psi_k,\psi_{k-1}}-1)A_{\psi_k,\psi_{k-1}}(t_k)\big]+O(\epsilon^2).
\end{multline}

Upon inserting Eq.(\ref{appmat2}) back to the matrix element (\ref{apppro}) and keeping terms only up to the first order in $|A_{\psi_i,\psi_{j}}|$ (which is small if $h(\tau)$
is slowly varying), we have:

\begin{multline}
 \langle x |S(\tau,0)| \psi_g \rangle_R = e^{-\int_0^{\tau} dt \lbrack E_{\psi_{g}}(t)+A_{\psi_g,\psi_g}(t)\rbrack }\cdot\\
\cdot \bigg( \langle x|\psi_g(\tau)\rangle_R-\sum\limits_{\psi\neq\psi_g}\langle x| \psi(\tau) \rangle_R \cdot\\
\cdot\int_0^{\tau}dt\;e^{-\int_t^{\tau}dt'\Delta E(t')}A_{\psi,\psi_g}(t)\bigg).
\end{multline}

Similarly, the other matrix element in Eq.(\ref{prob}), ${}_L\langle \psi_g(L_{\tau})|S(L_{\tau},\tau)|x\rangle$, reads:
\begin{multline}
 {}_L\langle \psi_g(L_{\tau})|S(L_{\tau},\tau)|x\rangle  =  e^{-\int_{\tau}^{L_{\tau}}dt\lbrack E_{\psi_g}(t)+A_{\psi_g,\psi_g}(t)\rbrack}\\
\bigg({}_L\langle \psi_g(\tau)|x\rangle-\sum\limits_{\psi\neq\psi_g}{}_L\langle\psi(\tau)|x\rangle\cdot\\
\cdot\int_{\tau}^{L_{\tau}}dt\;e^{-\int_{\tau}^tdt'\Delta E(t')}A_{\psi_g,\psi}(t)\bigg),
\end{multline}
where $\Delta E(t)\equiv E_{\psi}(t)-E_{\psi_g}(t)+A_{\psi,\psi}(t)-A_{\psi_g,\psi_{g}}(t)$.
Upon sending $L_x$ to infinity, the sum becomes an  integral over wavevectors $k$ and $\langle x|\psi_k(\tau)\rangle_R$ 
becomes the eigenstates tabulated in Eq.(\ref{appeigenr}) of the Appendix,
with $h=h(\tau)$.

Provided $h(\tau)<1$, to zeroth order the probability density is independent of $h$, $P(x,\tau)\simeq e^{-2|x|}$. 
The lowest order adiabatic correction to 
the probability density $\Delta P(x,\tau)=P(x,\tau)- P_o(\tau)$ reads: 
\begin{widetext}
\begin{multline}
\Delta P(x,\tau) = e^{-|x|-h(\tau)x} \int \frac{dk}{2\pi(1+c)} \psi_{{}_{L}k}(x;h(\tau)) 
\int_{\tau}^{L_{\tau}}dt e^{-\int_{\tau}^t dt'\Delta E(t')}A_{\psi_k,\psi_g}(t)-{}\\
-e^{-|x|+h(\tau)x}\int \frac{dk}{2\pi(1+c)}\psi_{{}_{R}k}(x;h(\tau))\int_{0}^{\tau}dt e^{-\int^{\tau}_t dt'\Delta E(t')}A_{\psi_k,\psi_g}(t),
\end{multline}
\end{widetext}
with $c=-\frac{1}{ik-h+1}$ (see Appendix \ref{sec:compdet}). $\int \Delta  P(x,\tau)dx=0$ and $P(x,\tau)$ is normalized to unity.

Upon substituting $x\rightarrow x-x_o(\tau)$ we can trivially tranform back to a ``moving defect'' 
frame of reference. In this frame, the zeroth order term reads $ P_o(\tau)=e^{-2|x-x_o(\tau)|}$.
\begin{figure}
\includegraphics[scale=0.7]{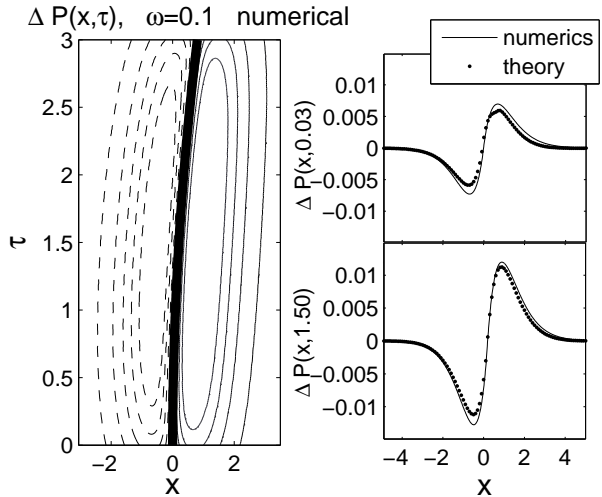}%
\caption{\label{fig:paperfigslow}Numerical solution $\Delta P(x,\tau)$  for a slowly moving defect. 
The thick solid line in the center represents the trajectory of the defect, $x_o(\tau)=18(1-\cos(\omega\tau))$, where $\omega=0.1$.
The time domain shown corresponds to slightly less than 5$\%$
of a complete period. The dashed and solid lines 
are respectively negative and positive contours of the adiabatic correction $\Delta P(x,\tau)$.
On the right: snapshots for times: $\tau=0.03$ and $\tau=1.50$ of $\Delta P(x,\tau)$ for both the analytic and the numerical solution. }
\end{figure}
\begin{figure}
\includegraphics[scale=0.62]{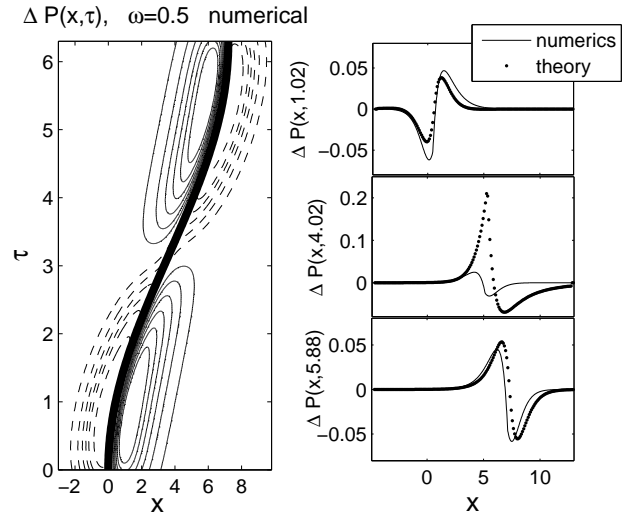}%
\caption{\label{fig:paperfigfast}Numerical solution $\Delta P(x,\tau)$  for a rapidly moving defect.
The thick solid line in the center represents a half-period of the defect trajectory, $x_o(\tau)=3.6(1-\cos(\omega\tau))$, where $\omega=0.5$. The dashed and solid lines 
are respectively negative and positive contours of the adiabatic correction $\Delta P(x,\tau)$.
On the right: snapshots for times: $1.02$, $4.02$ and $5.88$ of our approximate solution of $\Delta P(x,\tau)$ compared with the numerical result. }
\end{figure}

To check the accuracy of this result, we have used the moving defect frame and numerically integrated  the imaginary 
time Schroedinger equation for a slow sinusoidal defect trajectory $x_o(\tau)=18[1-\cos(0.1\tau)]$ with $L_{\tau}=3$
to obtain $\Delta P(x,\tau)=P(x,\tau)-e^{-2|x-x_o(\tau)|}$. The results are shown in Fig.\ \ref{fig:paperfigslow} \footnote{The corresponding 
tilt field for this trajectory is $h(\tau)=0.9\sin(\omega\tau)$, since with our rescaled units $h(\tau)=\frac{1}{2}\frac{dx_o}{d\tau}$.}. 
On the right hand side of Fig.\ref{fig:paperfigslow} we plotted two snapshots of the theoretical (lowest order correction)
and numerical $\Delta P(x,\tau)$ for two different times. The agreement between the two results is fairly good, even for 
$x$ and $\tau$ close to the boundaries. Note that the probability density piles up on the side of the radius of curvature, and is depleted on the opposite side.

However, the agreement deteriorates for rapidly changing defect trajectories. In Fig.\ \ref{fig:paperfigfast} we have plotted the numerically evaluated 
$\Delta P(x,\tau)$ for $x_o(\tau)=3.6[1-\cos(0.5\tau)]$ with $L_{\tau}\simeq 6.3$ and compared with the analytical result. The three snapshots on the right indicate 
significant deviation from the theory in the center of the range of $\tau$-values, where $\frac{dx_o}{d\tau}$ is changing rapidly. 
Note from the more exact numerical solution that probability again piles up on the side of the radius of curvature.  
 
\begin{acknowledgments}
We would like to thank M. Aziz, J. Blakely, G. Chan, V. Ignatesku and E. Williams for useful conversations. This work was supported by
the National Science Foundation through NSF Grant DMR-0231631 and the Harvard Materials Research Laboratory 
through NSF Grant NMR-0213805.
\end{acknowledgments}

\appendix*
\section{\label{sec:compdet}ANALYTICAL COMPUTATION OF THE PROBABILITY DENSITY}
In this Appendix we derive a number of results given in the main part of this paper. 
In most of what follows we assume that the lateral
dimension of the planar superconductor $L_x$ becomes infinite, although we shall be interested in the dependence of physical quantities on the
size of the \textit{timelike} dimension $L_{\tau}$. All the quantities are given in the dimensionless units discussed in Sec.\ref{sec:propag}.

\subsection{Eigenstates of the Hamiltonian}

The well known ground state of the Hamiltonian (\ref{hami}) with $h=0$ (in rescaled units) is simply $\psi_g(x)=e^{-|x|}$ with eigenenergy 
$E_g=-1$\footnote{See for example E. Merzbacher, (John Willey $\&$ Sons, New York, 1998), Chapter 6.}. The odd and even parity extended eigenstates
are respectively $\psi_k^{odd}(x)\sim\frac{1}{\sqrt{1+k^2}}(k\cos {kx} -\sin{k|x|})$ and $\psi^{even}_k(x)\sim\sin{kx}$ with $E_k=k^2$ and $k$ chosen to satisfy periodic boundary 
conditions in the $x$ direction.

This complete set of eigenstates can be used in the expansion (\ref{expa}), the sum being turned into an integral as  $L_x\to\infty$, which allows direct
computation of the propagator. Provided $|h|<1$, the case $h\neq 0$ can be reduced to $h=0$ via the ``imaginary gauge
transformation'' $e^{hx}H(h)e^{-hx}=H(0)$. In this case, Eq.(\ref{part3}) can be obtained by straightforward evaluation of the integral:
\begin{multline}
\mathcal{Z}_t[x,\tau;y,0;0]=e^{-|x|-|y|+\tau}+\frac{1}{\pi}\int_0^{\infty}dk\;e^{-k^2\tau}\cdot\\
\cdot\bigg(\frac{(k\cos kx-\sin k|x|)(k\cos ky-\sin k|y|)}{1+k^2}+\\
+\sin kx\sin ky\bigg).
\end{multline}
Above the delocalization transition, the nature of the eigenstates changes and there is no longer a bound state.
The non-Hermitian nature of the Hamiltonian (\ref{hami}) now becomes important. The right eigenstates for $h>0$ \footnote{The $h<0$ eigenstates 
can be derived by considering that $\mathcal{H}^{\dagger}(h)=\mathcal{H}(-h)$.} are (see Ref.\cite{hatan97}):
\begin{equation}\label{appeigenr}
\psi_{{}_{R}k}(x)\sim \left\{\begin{array}{ll}
e^{-ikx}+ce^{ikx-2hx} & \textrm{  for $x>0$}\\
(1+c)e^{-ikx}  & \textrm{  for $x<0$},
\end{array}\right.
\end{equation}
and the left ones:
\begin{equation}
\psi_{{}_{L}k}(x)\sim \left\{\begin{array}{ll}
(1+c)e^{ikx} & \textrm{  for $x>0$}\\
 e^{ikx}+ce^{-ikx+2hx} & \textrm{  for $x<0$}
\end{array}\right.
\end{equation}
with a common eigenenergy 
\begin{equation}
E_k=k^2+2ihk,
\end{equation} 
and
\begin{equation}
c=-\frac{1}{ik-h+1}.
\end{equation} 
The normalization condition is: $ \int dk |\psi_k\rangle_R \;{}_L\langle\psi_k|=1$.
Now the propagator is evaluated by computation of the integral $\int_{-\infty}^{\infty}dk \;e^{-k^2\tau}\psi_{{}_{L}k}(x)\psi_{{}_{R}k}(x)$. 
Remarkably, the final result for the propagator is again Eq.(\ref{part3}).

\subsection{Exact probability distribution of the fluctuating string}

In this subsection we sketch details for obtaining results used in Sec.\ref{sec:probabil}.

The $y$-integration in Eq.(\ref{probl}) can be computed analytically and is found to be:
\begin{widetext}
\begin{multline}\label{apinteg}
\int dy   \mathcal{Z}_t[y,L_{\tau};y',\tau_1;h]=1-\frac{e^{t-|y'|+hy'-h^2t}}{h^2-1}\mathrm{erfc}\left( \frac{|y'|}{2\sqrt{t}}-\sqrt{t}\right)-\\
-\frac{e^{hy'}}{2}\left(\frac{e^{h|y'|}}{1+h}\mathrm{erfc}\left( \frac{|y'|}{2\sqrt{t}}+h\sqrt{t}\right)+\frac{e^{-h|y'|}}{1-h} \mathrm{erfc}\left( \frac{|y'|}{2\sqrt{t}}-h\sqrt{t}\right)\right),
\end{multline}
\end{widetext}
where $t\equiv L_{\tau}-\tau_1$.

The $y$ integration in Eq.(\ref{problla}) is formally the same as the integral (\ref{apinteg}), $\tau_1$ and $y'$ having to be substituted with $\tau$ and $x-x_o(\tau)$ respectively.
The $y'$ integration is similarly found to be equal to:
\begin{multline}
\int dy' \mathcal{Z}_t[X,\tau;y',\tau_1;h]e^{-|y'|}=Q(h,X,\tau-\tau_1)+\\
+Q(-h,-X,\tau-\tau_1),
\end{multline}
where $X\equiv x-x_o(\tau)$ and
\begin{multline}
Q(h,x,t)=\frac{1}{2}e^{(1-2h) t-x}\mathrm{erfc}\left( \frac{-x}{2\sqrt{t}}+(1-h)\sqrt{t}\right)+\\
+\frac{e^{-h^2t-hx}}{2(2-h)}\bigg(e^{t-|x|}\mathrm{erfc}\left( \frac{|y'|}{2\sqrt{t}}-h\sqrt{t}\right)+\\
+e^{(1-h)^2t+(1-h)|x|}\mathrm{erfc}\left( \frac{|x|}{2\sqrt{t}}+(1-h)\sqrt{t}\right)\bigg).
\end{multline}


\begin{thebibliography}{14}
\expandafter\ifx\csname natexlab\endcsname\relax\def\natexlab#1{#1}\fi
\expandafter\ifx\csname bibnamefont\endcsname\relax
  \def\bibnamefont#1{#1}\fi
\expandafter\ifx\csname bibfnamefont\endcsname\relax
  \def\bibfnamefont#1{#1}\fi
\expandafter\ifx\csname citenamefont\endcsname\relax
  \def\citenamefont#1{#1}\fi
\expandafter\ifx\csname url\endcsname\relax
  \def\url#1{\texttt{#1}}\fi
\expandafter\ifx\csname urlprefix\endcsname\relax\def\urlprefix{URL }\fi
\providecommand{\bibinfo}[2]{#2}
\providecommand{\eprint}[2][]{\url{#2}}

\bibitem[{\citenamefont{Blatter et~al.}(1994)\citenamefont{Blatter, Feigel'man,
  Geshkenbein, Larkin, and Vinokur}}]{blatt94}
\bibinfo{author}{\bibfnamefont{G.}~\bibnamefont{Blatter}},
  \bibinfo{author}{\bibfnamefont{M.~V.} \bibnamefont{Feigel'man}},
  \bibinfo{author}{\bibfnamefont{V.~B.} \bibnamefont{Geshkenbein}},
  \bibinfo{author}{\bibfnamefont{A.~I.} \bibnamefont{Larkin}},
  \bibnamefont{and} \bibinfo{author}{\bibfnamefont{V.~M.}
  \bibnamefont{Vinokur}}, \bibinfo{journal}{Rev.\ Mod.\ Phys.}
  \textbf{\bibinfo{volume}{66}}, \bibinfo{pages}{1125} (\bibinfo{year}{1994}).

\bibitem[{\citenamefont{Hofstetter et~al.}(2004)\citenamefont{Hofstetter,
  Affleck, Nelson, and Schollwock}}]{hofstett04}
\bibinfo{author}{\bibfnamefont{W.}~\bibnamefont{Hofstetter}},
  \bibinfo{author}{\bibfnamefont{I.}~\bibnamefont{Affleck}},
  \bibinfo{author}{\bibfnamefont{D.~R.} \bibnamefont{Nelson}},
  \bibnamefont{and}
  \bibinfo{author}{\bibfnamefont{U.}~\bibnamefont{Schollwock}},
  \bibinfo{journal}{Europhys.\ Lett. B} \textbf{\bibinfo{volume}{66}},
  \bibinfo{pages}{178} (\bibinfo{year}{2004}).

\bibitem[{\citenamefont{Affleck et~al.}(2004)\citenamefont{Affleck, Hofstetter,
  Nelson, and Schollwock}}]{affl04}
\bibinfo{author}{\bibfnamefont{I.}~\bibnamefont{Affleck}},
  \bibinfo{author}{\bibfnamefont{W.}~\bibnamefont{Hofstetter}},
  \bibinfo{author}{\bibfnamefont{D.~R.} \bibnamefont{Nelson}},
  \bibnamefont{and}
  \bibinfo{author}{\bibfnamefont{U.}~\bibnamefont{Schollwock}},
  \bibinfo{journal}{J. Stat. Mech.} \textbf{\bibinfo{volume}{10}},
  \bibinfo{pages}{P10003} (\bibinfo{year}{2004}).

\bibitem[{\citenamefont{H.-C.Jeong and Williams}(1999)}]{jeong99}
\bibinfo{author}{\bibnamefont{H.-C.Jeong}} \bibnamefont{and}
  \bibinfo{author}{\bibfnamefont{E.~D.} \bibnamefont{Williams}},
  \bibinfo{journal}{Surf. Sc. Rep.} \textbf{\bibinfo{volume}{34}},
  \bibinfo{pages}{171} (\bibinfo{year}{1999}).

\bibitem[{\citenamefont{Hatano and Nelson}(1997)}]{hatan97}
\bibinfo{author}{\bibfnamefont{N.}~\bibnamefont{Hatano}} \bibnamefont{and}
  \bibinfo{author}{\bibfnamefont{D.~R.} \bibnamefont{Nelson}},
  \bibinfo{journal}{Phys.\ Rev. B} \textbf{\bibinfo{volume}{56}},
  \bibinfo{pages}{8651} (\bibinfo{year}{1997}).

\bibitem[{\citenamefont{Fisher et~al.}(1991)\citenamefont{Fisher, Fisher, and
  Huse}}]{fish91}
\bibinfo{author}{\bibfnamefont{D.~S.} \bibnamefont{Fisher}},
  \bibinfo{author}{\bibfnamefont{M.~P.~A.} \bibnamefont{Fisher}},
  \bibnamefont{and} \bibinfo{author}{\bibfnamefont{D.~A.} \bibnamefont{Huse}},
  \bibinfo{journal}{Phys.\ Rev. B} \textbf{\bibinfo{volume}{43}},
  \bibinfo{pages}{130} (\bibinfo{year}{1991}).

\bibitem[{\citenamefont{Hwa et~al.}(1993{\natexlab{a}})\citenamefont{Hwa,
  Nelson, and Vinokur}}]{hwa93b}
\bibinfo{author}{\bibfnamefont{T.}~\bibnamefont{Hwa}},
  \bibinfo{author}{\bibfnamefont{D.~R.} \bibnamefont{Nelson}},
  \bibnamefont{and} \bibinfo{author}{\bibfnamefont{V.~M.}~\bibnamefont{Vinokur}},
  \bibinfo{journal}{Phys. \ Rev. B} \textbf{\bibinfo{volume}{48}},
  \bibinfo{pages}{1167} (\bibinfo{year}{1993}{\natexlab{a}}).

\bibitem[{\citenamefont{Devereaux et~al.}(1994)\citenamefont{Devereaux,
  Scalettar, and Zimanyi}}]{derev94}
\bibinfo{author}{\bibfnamefont{T.~P.} \bibnamefont{Devereaux}},
  \bibinfo{author}{\bibfnamefont{R.~T.} \bibnamefont{Scalettar}},
  \bibnamefont{and} \bibinfo{author}{\bibfnamefont{G.~T.}
  \bibnamefont{Zimanyi}}, \bibinfo{journal}{Phys. \ Rev. B}
  \textbf{\bibinfo{volume}{50}}, \bibinfo{pages}{13625} (\bibinfo{year}{1994}).

\bibitem[{\citenamefont{Lee and Blakely}(1999)}]{blakely99}
\bibinfo{author}{\bibfnamefont{D.}~\bibnamefont{Lee}} \bibnamefont{and}
  \bibinfo{author}{\bibfnamefont{J.}~\bibnamefont{Blakely}},
  \bibinfo{journal}{Surf. Sc.} \textbf{\bibinfo{volume}{445}},
  \bibinfo{pages}{32} (\bibinfo{year}{1999}).

\bibitem[{\citenamefont{Hwa et~al.}(1993{\natexlab{b}})\citenamefont{Hwa,
  Le Doussal, Nelson, and Vinokur}}]{hwa93}
\bibinfo{author}{\bibfnamefont{T.}~\bibnamefont{Hwa}},
  \bibinfo{author}{\bibfnamefont{P.} \bibnamefont{Le Doussal}},
  \bibinfo{author}{\bibfnamefont{D.~R.} \bibnamefont{Nelson}},
  \bibnamefont{and} \bibinfo{author}{\bibfnamefont{V.~M.}~\bibnamefont{Vinokur}},
  \bibinfo{journal}{Phys. \ Rev. \ Lett.} \textbf{\bibinfo{volume}{71}},
  \bibinfo{pages}{3545} (\bibinfo{year}{1993}{\natexlab{b}}).

\bibitem[{\citenamefont{Doussal and Nelson}(1994)}]{dous94}
\bibinfo{author}{\bibfnamefont{P.~L.} \bibnamefont{Doussal}} \bibnamefont{and}
  \bibinfo{author}{\bibfnamefont{D.~R.} \bibnamefont{Nelson}},
  \bibinfo{journal}{Physica C} \textbf{\bibinfo{volume}{232}},
  \bibinfo{pages}{69} (\bibinfo{year}{1994}).

\bibitem[{\citenamefont{Feynman and Hibbs}(1965)}]{Feyn}
\bibinfo{author}{\bibfnamefont{R.~P.} \bibnamefont{Feynman}} \bibnamefont{and}
  \bibinfo{author}{\bibfnamefont{A.~R.} \bibnamefont{Hibbs}},
  \emph{\bibinfo{title}{Quantum Mechanics and path integrals}}
  (\bibinfo{publisher}{McGraw-Hill}, \bibinfo{year}{1965}).

\bibitem[{\citenamefont{Kleinert}(1995)}]{klein95}
\bibinfo{author}{\bibfnamefont{H.}~\bibnamefont{Kleinert}},
  \emph{\bibinfo{title}{Path Integrals in Quantum Mechanics, Statistics and
  Polymer Physics}} (\bibinfo{publisher}{World Scientific},
  \bibinfo{year}{1995}).

\bibitem[{\citenamefont{Mostafazadeh}(1997)}]{mostafa97}
\bibinfo{author}{\bibfnamefont{A.}~\bibnamefont{Mostafazadeh}},
  \bibinfo{journal}{Phys.\ Rev. A} \textbf{\bibinfo{volume}{55}},
  \bibinfo{pages}{1653} (\bibinfo{year}{1997}).

\end{thebibliography}
\end{document}